\documentclass[conference,compsoc]{IEEEtran}
\usepackage{booktabs}
\usepackage{caption}
\usepackage{multirow}
\usepackage{subcaption}
\usepackage{graphicx}
\usepackage{amssymb}
\usepackage{amsmath}
\usepackage{booktabs}
\usepackage{pifont}
\usepackage{color}
\usepackage{dblfloatfix}

\newcommand{\cmark}{\ding{51}}
\newcommand{\xmark}{\ding{55}}

\ifCLASSOPTIONcompsoc
  \usepackage[nocompress]{cite}
\else
  \usepackage{cite}
\fi

\ifCLASSINFOpdf
\else
\fi

\begin{document}

\title{A-Live: Passive Liveness Detection via Neuromuscular Micro-Motion Signatures on Commodity Sensors}


\author{\IEEEauthorblockN{Mohammed Gharib}
\IEEEauthorblockA{Aerendir Mobile Inc.\\
Mountain View, California, US, \\
Email: gharib@aermob.com}
\and
\IEEEauthorblockN{Sam Burns}
\IEEEauthorblockA{Aerendir Mobile Inc.\\
Mountain View, California, US, \\
Email: sam@aermob.com}
\and
\IEEEauthorblockN{Martin Zizi}
\IEEEauthorblockA{Aerendir Mobile Inc.\\
Mountain View, California, US, \\
Email: martin@aermob.com}}


\maketitle


\begin{abstract}
Liveness detection has evolved from a safeguard against presentation and replay attacks in biometric authentication to a broader requirement for distinguishing human users from non-human agents in modern digital systems. The emergence of generative and agentic AI further amplifies this need, positioning liveness as a fundamental security primitive.
Existing approaches face key limitations, including reliance on explicit user interaction, specialized hardware, vulnerability to increasingly realistic spoofing, and limited scalability in real-world deployments.
We present A-Live, a passive liveness detection framework that operates solely on inertial measurement unit (IMU) signals available in commodity devices. A-Live is based on the observation that neuromuscular micro-motions inherent to human motor control produce subtle but measurable signatures in inertial data, which are often treated as noise in prior work.
We design a lightweight feature extraction pipeline and a compact classifier suitable for real-time on-device deployment, and introduce a controllable physical micro-motion platform to evaluate robustness against engineered non-human motion.
Extensive evaluation across Android and iOS devices, including both automated and real-user settings, shows that A-Live achieves over 99.5\% accuracy with low false acceptance and rejection rates.
Our results demonstrate that neuromuscular micro-motion signatures provide a scalable and passive foundation for liveness detection under emerging AI-driven threat models.
\end{abstract}


%
\IEEEpeerreviewmaketitle

\section{Introduction}
\label{sec::intro}
Artificial intelligence (AI) has significantly increased the vulnerability of modern authentication systems to spoofing and impersonation attacks, making liveness detection a critical component of secure identity verification. In face recognition systems, for instance, the absence of liveness detection enables attackers to deceive authentication pipelines using high-fidelity 3D masks or reconstructed facial representations derived from publicly available images~\cite{usenix16liveness}.
More broadly, the emergence of generative and agentic AI systems has extended this problem beyond traditional biometric authentication. The core challenge is no longer limited to distinguishing real from fake biometric traits, but also to reliably differentiating human users from autonomous or semi-autonomous AI agents operating within digital systems. This distinction is increasingly relevant even in scenarios where such agents are authorized to act on behalf of users, as it introduces new requirements for accountability, authorization, and system-level trust.

This shift implies a broader security requirement in modern digital infrastructures, where human–agent interactions are becoming increasingly common. In domains such as financial services, e-commerce, and enterprise systems, actions may be initiated either directly by humans or delegated to automated agents, making reliable human presence verification an emerging security primitive in AI-driven environments.

Despite significant progress, existing liveness detection methods generally fall into three categories: challenge-based interaction methods, appearance-based visual analysis, and physiological or hardware-assisted sensing. While these approaches achieve high accuracy under controlled settings, they suffer from fundamental limitations, including user interaction friction, dependence on specialized hardware, susceptibility to increasingly realistic AI-driven attacks, and limited scalability in real-world deployments. More importantly, these approaches are typically evaluated under constrained settings that do not fully reflect the evolving threat model in which automated agents can perform large-scale, repeated authentication attempts. In such scenarios, even a low false acceptance rate can translate into a non-negligible number of successful compromises, exposing a gap between controlled evaluation performance and real-world security guarantees.

Beyond these evaluation limitations, recent studies have demonstrated successful attacks against state-of-the-art face recognition and liveness systems using low-cost spoofing strategies, highlighting the fragility of current assumptions when exposed to publicly available data and adaptive adversaries~\cite{sp23}. Importantly, emerging attack vectors are no longer limited to static spoofing artifacts, but increasingly involve actively generated motion patterns produced by robotic systems or AI-driven physical actuators that attempt to mimic human behavior. These attacks challenge a fundamental assumption underlying many existing approaches that motion consistent with human activity is sufficient to establish liveness. In practice, however, synthetic motion can be engineered to approximate coarse human dynamics, raising the question of whether finer-grained signals exist that are intrinsically tied to human physiology and thus resistant to such replication.

To address these limitations, we seek signals that are fundamentally rooted in human physiology rather than externally observable behavior. We find that evidence from neuroscience suggests that neural dynamics exhibit measurable signatures in peripheral motor activity, including cortico-muscular coupling and stable brain connectivity patterns~\cite{yale2015,cortico2015}. We hypothesize that neuromuscular micro-movements encode distinctive signatures of human motor control that are difficult to faithfully reproduce using synthetic or mechanically generated motion. Unlike coarse motion patterns, which can be approximated by programmable actuators, these micro-movements arise from complex neuromuscular processes and neural feedback loops, resulting in subtle, high-frequency variations that are inherently tied to human physiology. The prior work has largely treated involuntary micro-movements as noise and filtered them out to improve motion estimation accuracy~\cite{imuLiveness}. 

In this paper, we present A-Live, Aerendir-Liveness, a novel passive liveness detection framework that leverages IMU sensors available in commodity smartphones and wearable devices. A-Live captures subtle neuromuscular micro-movements reflected in inertial signals, which originate from human motor control processes influenced by brain activity. The proposed method operates in a fully passive manner, requiring no explicit user interaction or specialized hardware.
The name “Aerendir” is inspired by Eärendil the Mariner from J.R.R. Tolkien’s legendarium, symbolizing guidance through uncertainty. Analogously, A-Live aims to provide a trust layer for secure human verification in environments populated by autonomous agents and adversarial systems.
We design a lightweight feature extraction pipeline and a compact classifier suitable for real-time deployment on resource-constrained devices. Since IMU sensors are already embedded in modern smartphones and wearables, the system is inherently scalable and widely deployable.

We conduct extensive experimental evaluation across diverse real-world and adversarial settings. Our study includes live interactions with genuine human users, as well as large-scale automated attack scenarios executed on device farms spanning 101 smartphone and tablet models, covering the majority of commercially available devices across both Android and iOS platforms. To further assess robustness against active physical attacks, we design and implement a controllable robotic motion platform capable of generating a wide range of synthetic movement patterns that approximate human behavior.
Across these settings, A-Live achieves over 99.5\% detection accuracy while maintaining consistently low false acceptance and false rejection rates.

The contributions of this paper are threefold:
\begin{itemize}
\item We introduce a novel signal modality for liveness detection based on neuromuscular micro-movements captured by commodity IMU sensors;
\item We implement a lightweight, fully passive system that achieves high accuracy across real-world, cross-device deployments;
\item We demonstrate robustness against both passive spoofing and actively generated synthetic motion, including attacks implemented using a controllable robotic motion platform designed to mimic human-like dynamics.
\end{itemize}

Together, these contributions highlight the limitations of existing liveness detection paradigms and demonstrate the potential of neuromuscular micro-motion as a fundamentally different and more robust signal. 
We next review prior work in liveness detection and related sensing approaches in Section~\ref{sec::relatedWork}, followed by physiological motivation in Section~\ref{sec::physiology}, which motivates our problem formulation and adversarial model in Section~\ref{sec::problem}. 
The proposed A-Live framework is then introduced in Section~\ref{sec::proposal}, followed by experimental evaluation in Section~\ref{sec::evaluation} and a discussion of implications and future work in Section~\ref{sec::conclusion}.

\section{Related Work}
\label{sec::relatedWork}
Liveness detection has been widely studied, essentially to defend biometric systems against presentation and replay attacks. Existing approaches can be broadly categorized into three groups: (i) challenge-based, mainly active methods, (ii) appearance-based, mainly passive methods, and (iii) physiological signal-based methods. We review each category and highlight their limitations in the context of robust biometric liveness detection.

\textbf{Challenge-based methods} require explicit user interaction and verify the correctness of a prompted response. Common examples include facial gesture-based schemes (e.g., blinking~\cite{Wu2015}, smiling~\cite{void}, or head rotation), and voice-based challenge--response systems. More advanced approaches combine visual input with additional sensing modalities. For instance, \cite{imuLiveness} proposes correlating facial motion observed by the camera with inertial measurements obtained from the device while the user performs controlled movements. Related mechanisms such as CAPTCHA systems~\cite{captcha} also follow a challenge--response paradigm, but do not constitute biometric liveness detection since they do not verify physiological authenticity. While effective against simple replay attacks, challenge-based methods suffer from several limitations. They introduce user friction, require explicit cooperation, and may degrade usability. Moreover, they remain vulnerable to sophisticated attacks, such as video replay or mechanically assisted responses.

\textbf{Appearance-based methods} operate passively by analyzing visual cues without requiring user interaction. Early work focused on texture-based anti-spoofing, exploiting differences between real faces and printed or displayed images~\cite{maata}. Recent advances leverage deep learning to learn discriminative features for presentation attack detection~\cite{Liu2018,Zhang2020}. Other approaches utilize reflection and illumination analysis~\cite{Tan2010}, or depth information obtained from stereo or structured light sensors~\cite{Atoum2017}. 
A related line of work explores remote photoplethysmography (rPPG), which estimates subtle color variations caused by blood flow from facial videos. Existing methods can be broadly categorized into color variation–based approaches~\cite{Verkruysse2008} and blind source separation (BSS)-based approaches~\cite{McDuff2010}. While these methods capture physiological signals in a non-contact manner, they primarily rely on periodic pulse estimation and operate on visual data without explicitly modeling underlying biological processes. Hence, they are sensitive to lighting conditions, motion artifacts, and camera quality. Overall, appearance-based approaches are attractive due to their low deployment cost and passive operation. However, they predominantly rely on surface-level cues and can be vulnerable to high-quality presentation attacks, including masks, high-resolution displays, and deepfake-based forgeries \cite{sp23,imuSP23}.

\textbf{Physiological signal-based methods} aim to detect liveness by exploiting intrinsic biological signals associated with human physiology. In this category, a large body of work utilizes \textit{contact-based or specialized measurements} such as electrocardiography (ECG) \cite{Odinaka2012ECG}, electromyography (EMG) \cite{emg}, fingerprint pulse or blood-flow based liveness detection~\cite{livenessFinger}, as well as iris-based biometric systems, including recent large-scale commercial iris-based deployments~\cite{worldcoinWhitePaper}. These approaches provide strong guarantees of liveness and uniqueness due to the inherent difficulty of spoofing physiological signals. However, they rely on dedicated or contact-based hardware (e.g., the Orb for high-quality iris acquisition), often requiring users to be physically present at specific locations. This dependence introduces significant scalability challenges and user friction, limiting their practicality in large-scale and ubiquitous deployment scenarios. 

\textit{Non-contact physiological sensing} approaches have also been extensively studied, including ballistocardiography (BCG)~\cite{bcg}, and camera-based iris liveness detection~\cite{irisLiveness}. These approaches leverage commodity imaging devices, improving accessibility and ease of deployment. However, this increased practicality typically comes at the cost of reduced robustness, as such methods are more susceptible to noise, environmental variations, and presentation attacks. In addition, they often require explicit user cooperation and introduce non-negligible interaction overhead, which can further impact usability in real-world applications.

Despite significant progress, existing liveness detection methods exhibit a fundamental trade-off between usability, robustness, and deployment scalability. Challenge-based approaches introduce user friction and remain vulnerable to replay or assisted attacks. Appearance-based methods, while passive and easy to deploy, rely on surface-level cues and are susceptible to high-quality forgeries and environmental variations. Physiological approaches offer stronger guarantees of liveness, but often depend on specialized hardware or constrained sensing conditions, limiting their practicality at scale. A key limitation across these categories is the lack of methods that simultaneously achieve (i) passive operation, (ii) robustness to sophisticated automated and physical attacks, and (iii) scalability across commodity devices without specialized hardware.

It is worth noting that prior work has explored the use of IMU sensors for liveness detection and user behavior analysis~\cite{imuLiveness}. However, these approaches primarily focus on coarse-grained motion patterns and explicitly suppress fine-grained involuntary variations as noise through filtering and smoothing. In contrast, we hypothesize that such micro-movements encode intrinsic physiological signals that are difficult to replicate by automated agents or mechanical systems. This shift in perspective forms the basis of the physiological modeling introduced in the following section.

\section{Physiological Motivation}
\label{sec::physiology}

Recent developments in biometric and authentication systems have explored physiological and behavioral signals for establishing robust notions of identity and liveness, with early foundations of this direction articulated in prior work on physiological-based authentication and liveness estimation~\cite{livenessPatent}. These systems are motivated by the observation that biological processes introduce structured variability that is difficult to reproduce using synthetic or automated agents.

Complementary findings in neuroscience suggest that both central and peripheral physiological signals carry individual-specific structure. For example, stable functional connectivity patterns in brain activity, often referred to as \textit{connectome fingerprints}, have been demonstrated using functional magnetic resonance imaging (fMRI), a high-cost and specialized neuroimaging modality, indicating that large-scale neural organization exhibits intrinsic individuality~\cite{yale2015}. In addition, motor neuroscience studies have shown that cortical activity is functionally coupled to muscular responses through the corticospinal system, where neural activation patterns are translated into coordinated muscle activity during voluntary and semi-involuntary motion~\cite{cortico2015}.

From a physiological perspective, motor commands emerge as distributed neural activations across populations of motor units rather than single synchronized oscillations. These signals propagate through the peripheral nervous system to multiple muscle groups, producing subtle, continuous micro-variations even during low-motion or near-static conditions. Such variations reflect underlying neural control processes rather than purely mechanical noise.

While prior evidence of neural individuality relies on medical-grade imaging modalities such as fMRI, we investigate whether peripheral manifestations of these processes can be observed using commodity sensing hardware. Although cortical activity is not directly observable in practical authentication settings, its downstream effects may be captured indirectly through high-sensitivity inertial sensors. This motivates the hypothesis that fine-grained micro-movements measured via commodity IMU sensors encode structured signatures of neuromuscular control, and can therefore serve as a discriminative signal for liveness detection beyond conventional behavioral or appearance-based cues. 

More broadly, this perspective suggests that liveness can be formulated as a fundamental security primitive grounded in observable neurophysiological side channels, rather than as a task-specific biometric verification problem, motivating the design of a practical system that operationalizes this signal in real-world sensing conditions.

\section{Problem Formulation and Adversarial Model}
\label{sec::problem}

In this section, we formalize the problem addressed in this work and describe the system and adversarial assumptions under which our method operates. We organize the discussion into two parts: (i) the system model and problem formulation, which define liveness detection as a binary classification problem over inertial sensor time-series signals and specify the sensing capabilities and operational constraints of commodity devices, and (ii) the adversarial setting, which characterizes the capabilities and limitations of non-human agents in generating interaction traces that may attempt to mimic human-induced sensor patterns. We further note that we use the terms \textit{human} and \textit{live} interchangeably, and \textit{non-human} and \textit{non-live} interchangeably, throughout the paper.

\subsection{System Model and Problem Formulation}

We consider the problem of distinguishing whether an interaction in a sensing-enabled system originates from a physically present human user or from a non-human source such as a software agent or mechanically induced process. Unlike traditional biometric authentication, which focuses on identity verification, our objective is to establish the presence of a live human based on intrinsic motion characteristics.
Crucially, we focus on fine-grained motion dynamics that reflect underlying neuromuscular activity, rather than coarse behavioral patterns. These micro-scale signals are inherently constrained by human physiology and are therefore significantly harder to reproduce using synthetic or automated means, forming the basis for robust liveness inference in adversarial settings.

We consider a standard mobile and wearable sensing environment in which IMU sensors are widely available on commodity devices, including smartphones and wearable platforms. These sensors provide continuous streams of accelerometer and gyroscope measurements that reflect device motion in real-world usage settings.
The system operates passively, without requiring explicit user interaction or user cooperation, and relies solely on sensor data that can be accessed within standard application execution environments, including mobile and browser-based contexts where such interfaces are exposed.
Sensor readings are assumed to capture a mixture of voluntary user motion, involuntary neuromuscular activity, and environmental or device-induced perturbations.

Formally, let $\mathbf{X} = \{x_t\}_{t=1}^{N}$ denote a multivariate time-series signal captured from commodity IMU sensors during an interaction window, where $x_t \in \mathbb{R}^d$ represents the IMU measurement vector at sample index $t$, consisting of accelerometer and gyroscope readings across three axes, where $d = 6$. The goal is to learn a function $f(\mathbf{X}) \rightarrow \{0,1\}$ that maps the observed sequence to a binary decision indicating human presence $(1)$ or non-human origin $(0)$, without requiring explicit user effort or specialized hardware.
Human interactions exhibit structured yet stochastic micro-movement patterns governed by neuromuscular control processes, while non-human interaction traces are generated through algorithmic, synthetic, or mechanically induced motion mechanisms. Such non-human signals may reproduce coarse-grained motion characteristics, but are not constrained by the same physiological control dynamics that govern human motor behavior.

\subsection{Adversarial Setting}

The adversary in our setting aims to generate interaction traces that are indistinguishable from those produced by a legitimate human user. This may include automated scripts, software agents, or AI-driven systems capable of simulating user interactions at the input level.

We assume a strong adversary capable of observing system behavior and generating interaction traces using software agents, AI-driven control policies, or mechanically actuated systems. Such an adversary may attempt to replicate both coarse and fine grained motion characteristics of legitimate users.
In addition, we explicitly consider physical adversaries that can induce motion on the device through robotic mechanisms or external actuation, enabling a broad class of engineered motion patterns designed to mimic human interaction dynamics.
Despite these capabilities, adversaries remain fundamentally constrained by the lack of access to neuromuscular control processes that generate biologically induced micro-movements in genuine human interactions. As a result, faithfully reproducing the statistical structure of these micro-movements remains challenging even under highly controlled physical spoofing conditions.

In practical deployments, IMU data is typically transmitted through secure system pipelines (e.g., operating system sensor frameworks and trusted execution environments), which limit adversarial manipulation of raw sensor streams. We do not rely on these protections for correctness; instead, our goal is to provide a sensing-layer separation between biologically grounded and synthetically or mechanically generated interaction traces.
The core challenge addressed in this work is therefore the reliable discrimination of human-generated IMU motion traces from both software-driven and physically synthesized alternatives under realistic sensing conditions, without restricting the adversary to any specific attack modality.

\section{ Proposed Method: A-Live}
\label{sec::proposal}

\begin{figure*}[!t]
    \centering         
    \includegraphics[width=\linewidth]{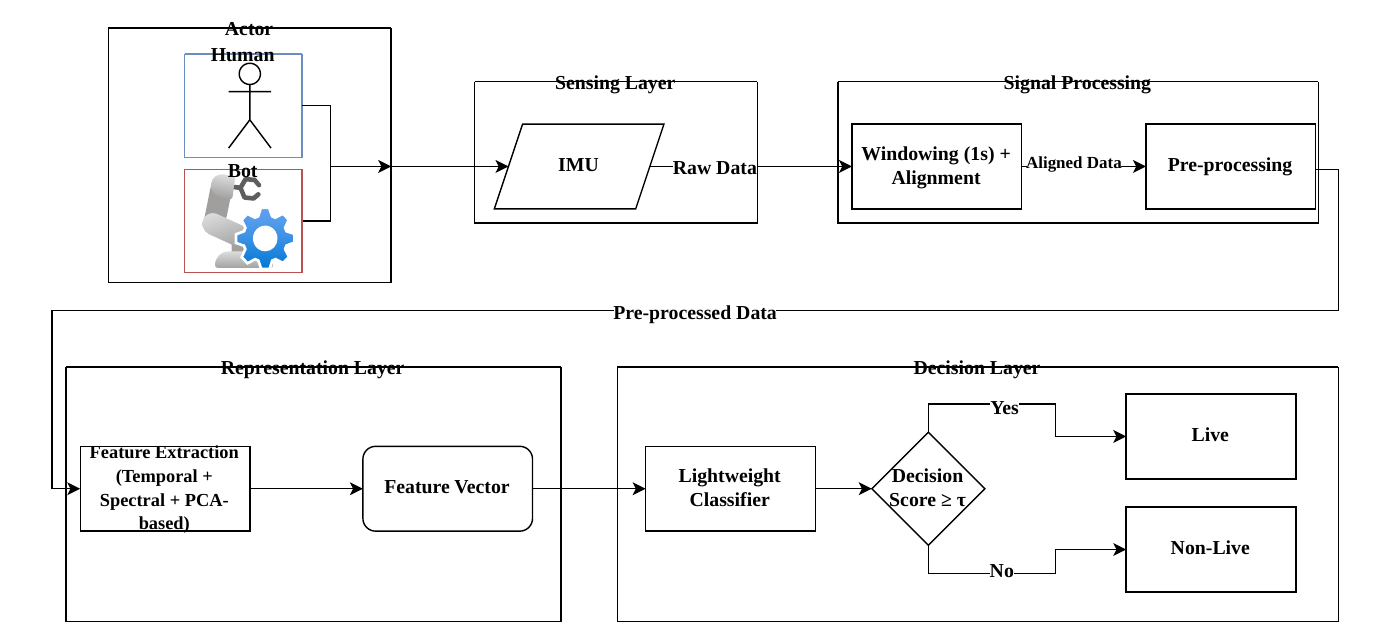}
    \caption{Overview of the A-Live Liveness Detection Framework }
    \label{fig:pipeline}
\end{figure*}

We present \textit{A-Live}, a liveness detection mechanism based on the observation that involuntary micro-movements driven by human motor control induce distinctive stochastic signatures in inertial sensor measurements that are difficult to reproduce using synthetic or non-biological motion processes.
A-Live operates as a five-stage pipeline illustrated in Figure~\ref{fig:pipeline}. The system begins with an interaction actor, which may correspond to either a legitimate human user or an adversarial agent attempting to generate spoofed interaction traces. The interaction is captured through commodity IMU sensors, producing raw motion signals. These signals are then processed in a signal processing stage involving windowing, temporal alignment, and pre-processing. Next, a representation layer transforms the processed signals into discriminative feature vectors through temporal, spectral, and statistical feature extraction. Finally, a lightweight classifier operates on the extracted representation to produce a binary decision indicating live or non-live interaction.

The remainder of this section is organized as follows. We first describe the signal Acquisition and pre-processing stage, followed by the feature extraction process that captures temporal, spectral, and stochastic motion characteristics, and finally present the classification stage that produces the binary live versus non-live decision.
To support the mathematical formulation of the system, Table~\ref{tab:notation} summarizes the notation used throughout the paper.

\begin{table}[!t]
\centering
\caption{Summary of Notation Used in A-Live}
\label{tab:notation}
\begin{tabular}{ll}
\toprule
\textbf{Symbol} & \textbf{Description} \\
\midrule
$T$ 
& Window duration (seconds) \\
$f_s$ 
& Sampling frequency \\
$N = T f_s$ 
& Number of samples per window \\
$\mathbf{X} = \{x_t\}_{t=1}^{N}$ 
& IMU time-series within a window of duration $T$ \\
$x_t \in \mathbb{R}^d$ 
& IMU measurement vector at sample index $t$ \\
$\mathbf{X}_k$ 
& $k$-th windowed segment of the signal \\
$\mathbf{M}(t)$ 
& Voluntary (macro) motion component \\
$\mathbf{N}(t)$ 
& Neuromuscular micro-movement component \\
$\mathbf{E}(t)$ 
& Environmental noise and sensor artifacts \\
$\phi(\cdot)$ 
& Feature extraction mapping function \\
$\mathbf{z}_k = \phi(\mathbf{X}_k)$ 
& Feature vector for window $\mathbf{X}_k$ \\
$g(\mathbf{z}_k)$ 
& Classifier score (real-valued output) \\
$f(\mathbf{z}_k) \in \{0,1\}$ 
& Binary decision function (live / non-live) \\
$\tau$ 
& Decision threshold \\
\bottomrule
\end{tabular}
\end{table}

\subsection{Signal Acquisition and Pre-processing}
\label{sec:signal}

This subsection describes the signal acquisition pipeline and pre-processing strategy used by A-Live. We first outline the IMU sensing modality and data structure, followed by practical considerations including sampling variability and sensor misalignment. We then introduce the signal refinement design principles, which aim to preserve fine-grained motion structure while improving robustness.

\textbf{IMU Signal Acquisition.}
A-Live operates on IMU data collected from commodity mobile devices. Specifically, we utilize tri-axial accelerometer and gyroscope sensors, which measure linear acceleration (in $\mathrm{m/s^2}$) and angular velocity (in $\mathrm{rad/s}$), respectively. Each sensor produces a time series of samples of the form $(t, x, y, z)$, where $t$ denotes the timestamp and $(x, y, z)$ correspond to measurements along the three spatial axes.

The sampling rate of IMU sensors varies across devices and operating systems, typically ranging from tens to several hundreds of Hertz (e.g., $\sim 50-1200\,\mathrm{Hz}$ in common smartphone configurations). To improve robustness across heterogeneous hardware configurations, A-Live processes fixed-duration windows of one second, regardless of the underlying sampling frequency. This design allows the method to remain agnostic to device-specific sampling configurations while preserving sufficient temporal resolution for capturing micro-movement dynamics. 

The continuous signal is segmented into fixed-duration windows for analysis. Let $T$ denote the window duration (here, $T = 1\,\mathrm{s}$) and $f_s$ the sampling frequency, such that each window contains $N = T f_s$ samples. We define the $k$-th windowed segment as:
\begin{equation}
\mathbf{X}_k = \{ x_t \mid t \in [kN, (k+1)N) \},
\end{equation}
where $\mathbf{X}_k$ represents the $k$-th temporal segment used for feature extraction and classification.

The recorded signals capture a superposition of multiple components: (i) voluntary, coarse-grained motion induced by user interaction (e.g., touch or device handling), (ii) fine-grained micro-movements arising from neuromuscular activity, and (iii) sensor and environmental noise. A-Live is designed to exploit statistical signatures associated with the neuromuscular micro-movement component within the observed motion stream.

\begin{figure*}
         \begin{subfigure}{0.333\textwidth}
         \centering         \includegraphics[width=\textwidth]{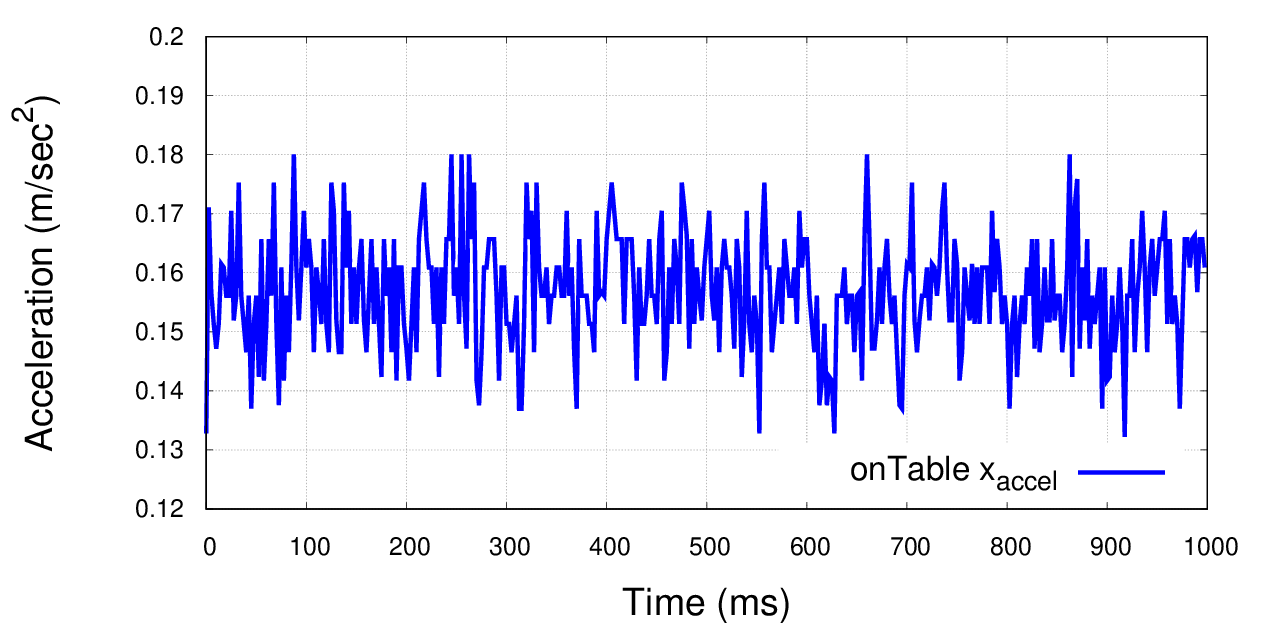}
         \caption{ On Table $(x_{accel})$}
         \label{fig:onTable_accel}
     \end{subfigure}
     \begin{subfigure}{0.333\textwidth}
         \centering         \includegraphics[width=\textwidth]{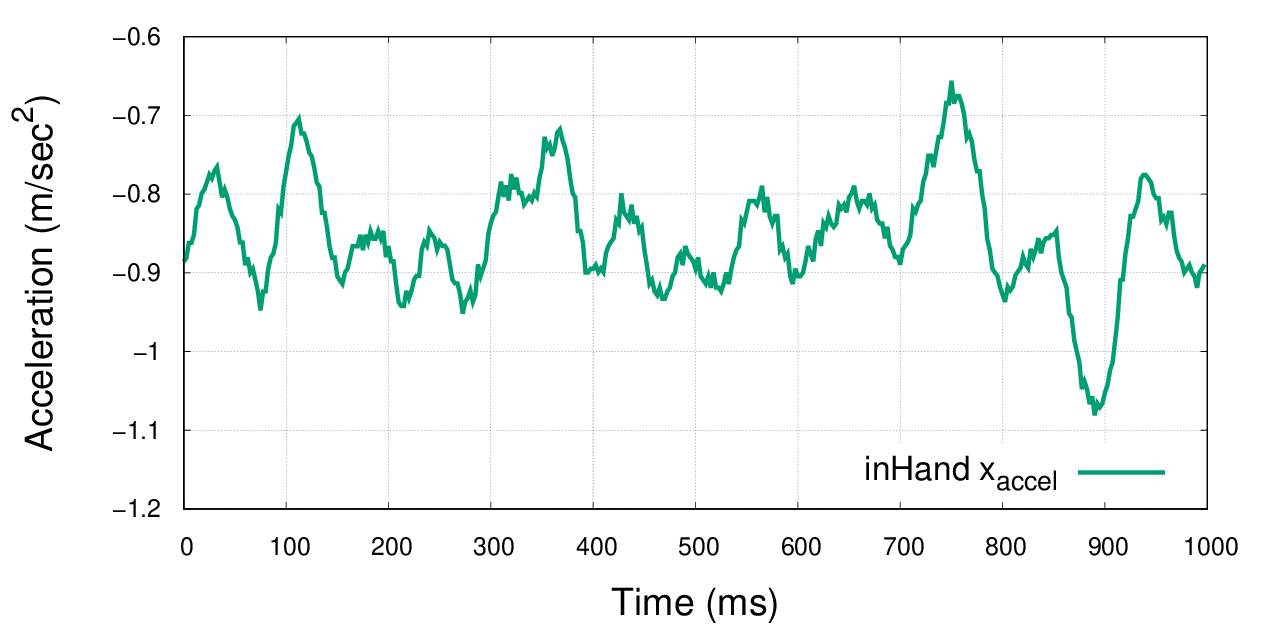}
         \caption{ In Hand $(x_{accel})$}
         \label{fig:inHand_accel}
     \end{subfigure}
     \begin{subfigure}{0.333\textwidth}
         \centering         \includegraphics[width=\textwidth]{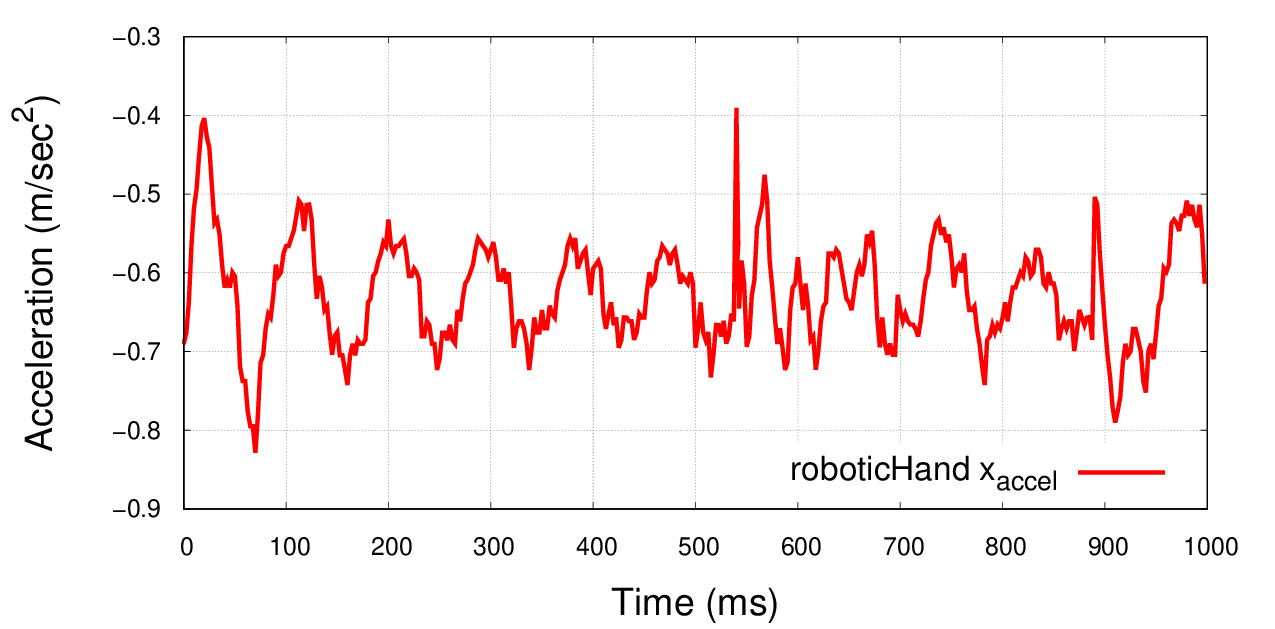}
         \caption{On Robotic Hand $(x_{accel})$}
         \label{fig:roboticHand_accel}
     \end{subfigure}\\
     
     \begin{subfigure}{0.333\textwidth}
         \centering         \includegraphics[width=\textwidth]{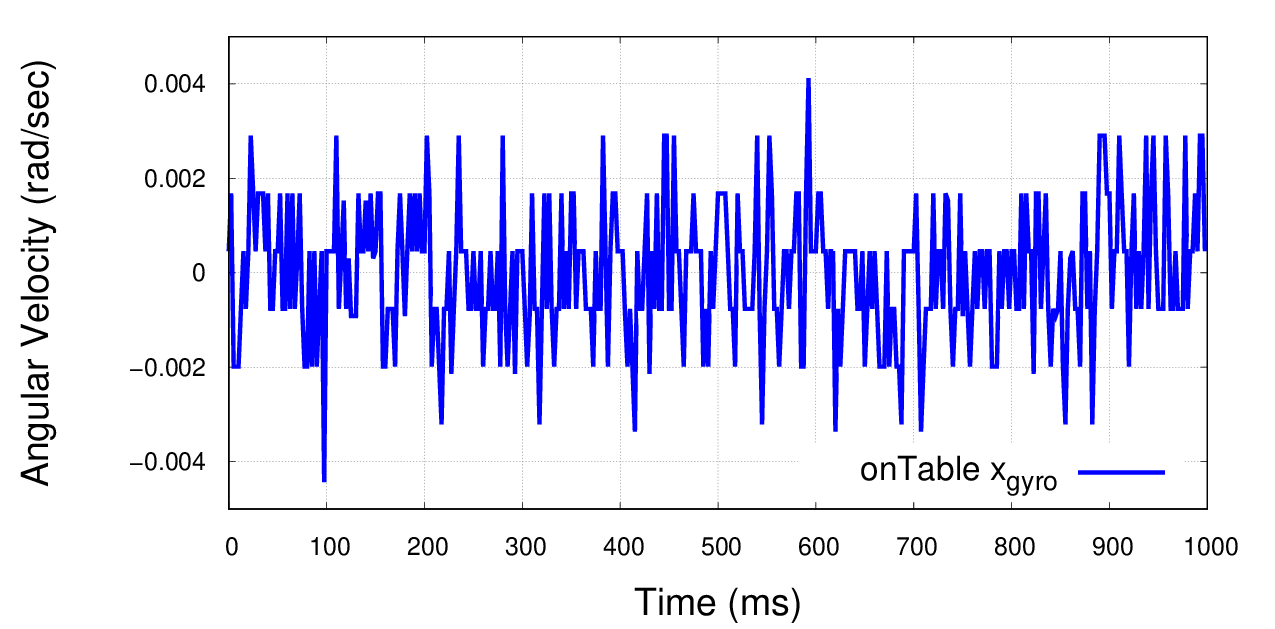}
         \caption{ On Table $(x_{gyro})$}
         \label{fig:onTable_gyro}
     \end{subfigure}
     \begin{subfigure}{0.333\textwidth}
         \centering         \includegraphics[width=\textwidth]{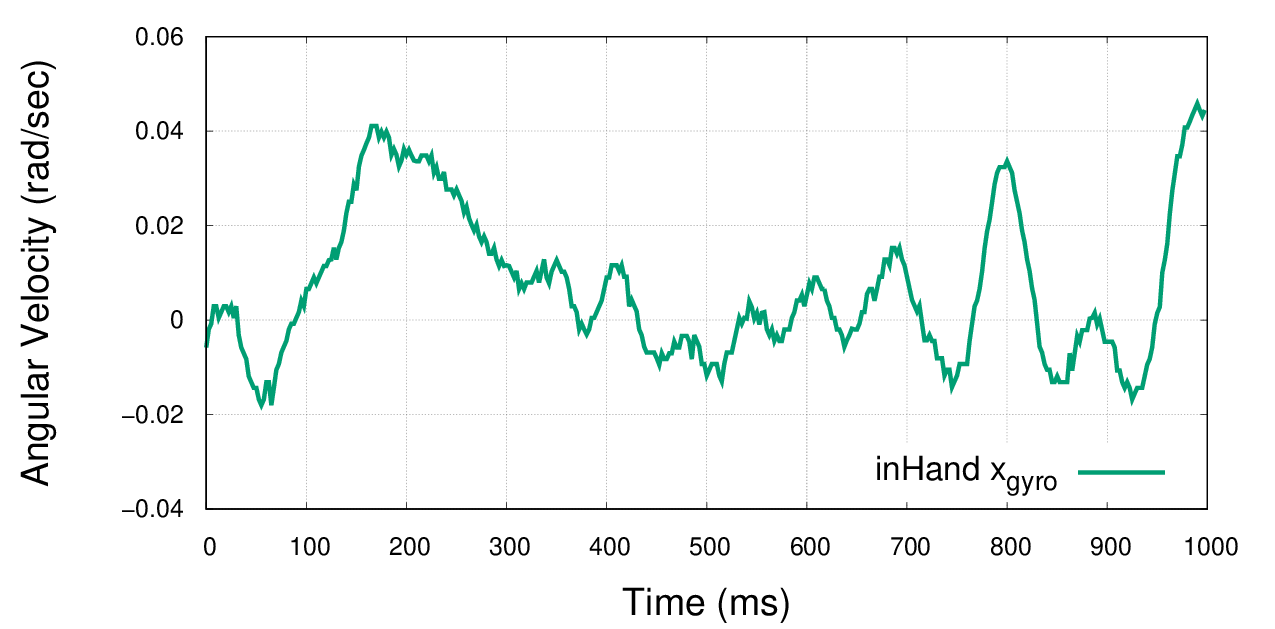}
         \caption{ In Hand $(x_{gyro})$}
         \label{fig:inHand_gyro}
     \end{subfigure}
     \begin{subfigure}{0.333\textwidth}
         \centering         \includegraphics[width=\textwidth]{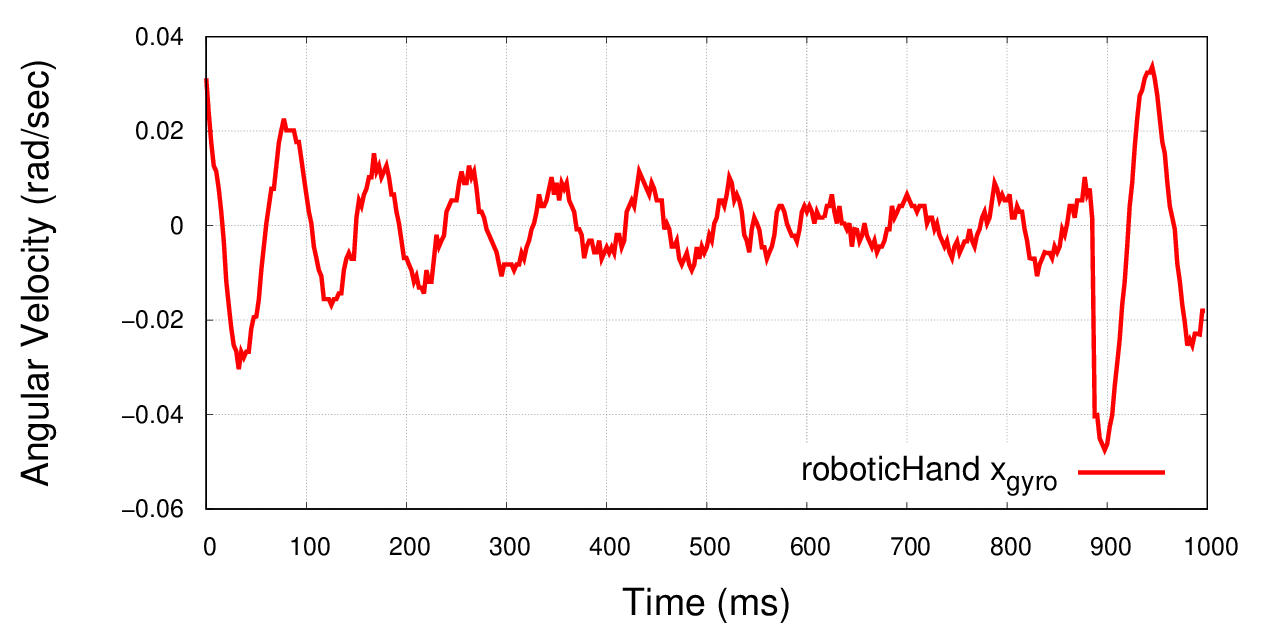}
         \caption{On Robotic Hand $(x_{gyro})$}
         \label{fig:roboticHand_gyro}
     \end{subfigure}
        \caption{IMU Sensor Traces Under On-Table, In-Hand, and Robotic Micro-Motion Conditions }
        \label{fig:rawdata}
\end{figure*}

We model the observed inertial signal as a superposition of multiple underlying components:
\begin{equation}
\mathbf{X}(t) = \mathbf{M}(t) + \mathbf{N}(t) + \mathbf{E}(t),
\end{equation}
where $\mathbf{M}(t)$ represents voluntary, coarse-grained motion induced by user interaction, $\mathbf{N}(t)$ denotes fine-grained neuromuscular micro-movements, and $\mathbf{E}(t)$ captures environmental perturbations and sensor noise. 
The key premise of A-Live is that $\mathbf{N}(t)$ contains structured stochastic patterns characteristic of human motor control, which are difficult to reproduce through synthetic or mechanically generated motion. These components manifest differently across interaction contexts, resulting in distinct inertial signatures depending on whether the device is stationary, held by a user, or mechanically actuated. Figure~\ref{fig:rawdata} illustrates representative accelerometer and gyroscope traces under these three conditions, highlighting the structural differences between natural human interaction, passive device placement, and mechanically generated motion representing physical adversarial attempts to emulate human interaction dynamics. For clarity of presentation, we show only the $x$-axis components of the accelerometer and gyroscope signals, as all three axes exhibit consistent qualitative trends across the different conditions.

\textbf{Temporal Misalignment.}
A practical challenge in IMU-based sensing is the lack of strict temporal synchronization between accelerometer and gyroscope streams which is also noticed by recent work \cite{imuGlitch}. Such misalignment can introduce subtle inconsistencies in multimodal sensor fusion and can even be exploited in adversarial settings~\cite{imuGlitch}. In our setting, this manifests as small but non-negligible timestamp discrepancies between corresponding sensor samples. A-Live accounts for this effect during pre-processing by applying alignment strategies that mitigate cross-sensor temporal skew before downstream analysis.

\textbf{Signal Refinement.}
The raw IMU signals are first conditioned to improve robustness while preserving the intrinsic structure of the observed motion. Rather than performing aggressive filtering or attempting explicit signal separation, A-Live adopts a conservative pre-processing strategy that attenuates dominant nuisance components while retaining structurally informative variations in the signal that may encode liveness cues. In particular, this stage aims to: (i) suppress slow-varying trends induced by device drift and large-scale motion (e.g., sustained hand movement or intentional shaking), (ii) reduce measurement noise and sensor-specific artifacts without enforcing strict frequency band separation, and (iii) preserve the intrinsic multi-scale temporal structure of the signal for downstream analysis.

A fundamental challenge in IMU-based analysis is that biologically induced micro-movements, voluntary motion, and environmental noise often overlap in both amplitude and frequency domains. As a result, precise separation of these components through pre-processing alone is inherently unreliable and risks removing liveness-relevant signal characteristics. In particular, neuromuscular micro-movements remain embedded within broader motion patterns and cannot be cleanly isolated using simple filtering criteria.

Accordingly, A-Live does not attempt to explicitly disentangle signal components at the pre-processing stage. Instead, it retains a structurally rich representation of the motion signal and defers fine-grained discrimination to the feature extraction stage, where informative representations are constructed to capture neuromuscular micro-movement signatures while suppressing irrelevant variations. This design enables the subsequent lightweight classifier to operate on a compact and discriminative feature space, rather than raw or minimally processed signals.

The filtering operations in this stage are therefore limited to signal conditioning for robustness and are intentionally distinct from the frequency-selective analysis performed during feature extraction. In particular, the feature extraction stage performs structured analysis across multiple temporal and spectral scales to isolate liveness-relevant characteristics, whereas the present stage avoids aggressive transformations to preserve information for downstream representation learning.

\subsection{A-Live Feature Extraction}
\label{sec:features}
{
We describe the feature extraction stage of A-Live at a high level, focusing on the underlying design principles rather than implementation-specific details. The objective is to transform pre-processed inertial sensor signals into a compact representation that captures discriminative structure in fine-grained motion dynamics, while maintaining robustness to device heterogeneity and environmental variability. 
To support reproducibility and independent validation, without exposing
proprietary implementation details, we provide a reference implementation that demonstrates the full data acquisition and inference pipeline~\cite{aerendirA_Live}.

Given a pre-processed signal window $\mathbf{X}_k$, A-Live computes a feature representation via a mapping function:
\begin{equation}
\mathbf{z}_k = \phi(\mathbf{X}_k),
\end{equation}
where $\phi(\cdot)$ denotes the feature extraction pipeline. 
This mapping aggregates multiple complementary descriptors, which can be conceptually grouped as:
\begin{equation}
\phi(\mathbf{X}_k) = \left[
\phi_{\text{temp}}(\mathbf{X}_k),\;
\phi_{\text{spec}}(\mathbf{X}_k),\;
\phi_{\text{stat}}(\mathbf{X}_k)
\right],
\end{equation}
corresponding to temporal, spectral, and stochastic feature families, respectively.

\textbf{Multi-Stage Feature Construction.}
A-Live constructs features from tri-axial accelerometer and gyroscope signals using multiple representations of the same signal window. Descriptors are computed both in the original axis-aligned space and in a decorrelated subspace obtained via Principal Component Analysis (PCA), improving robustness to device orientation and axis coupling effects.
The pipeline first validates each window to reject degenerate cases (e.g., static or excessively noisy segments). The signal is then transformed into complementary representations, including time-domain signals, frequency-domain decompositions, and normalized magnitude-based forms.

These feature families are instantiated through a combination of descriptors computed over both axis-aligned signals and decorrelated PCA-based representations.
On these representations, A-Live extracts descriptors capturing distinct aspects of motion behavior:
\begin{itemize}
\item \textit{Temporal variability features} describing the evolution and smoothness of motion trajectories.
\item \textit{Spectral structure features} characterizing the distribution of signal energy across frequency components.
\item \textit{Stochastic and distributional features} quantifying irregularity, impulsiveness, and structural complexity.
\end{itemize}

While the full feature set is high-dimensional and operates in a joint representation space that is not directly visualizable, it is useful to illustrate the effect of the engineered feature design in practice.
To this end, Figure~\ref{fig:features} presents representative features that illustrate the behavior of the feature extraction stage.
Specifically, we show the temporal evolution of sample entropy, spectral entropy, and fractal complexity computed over successive signal segments. These illustrative features are representative but do not exactly correspond to the full feature set used by the classifier. 

\begin{figure*}
         \begin{subfigure}{0.333\textwidth}
         \centering         \includegraphics[width=\textwidth]{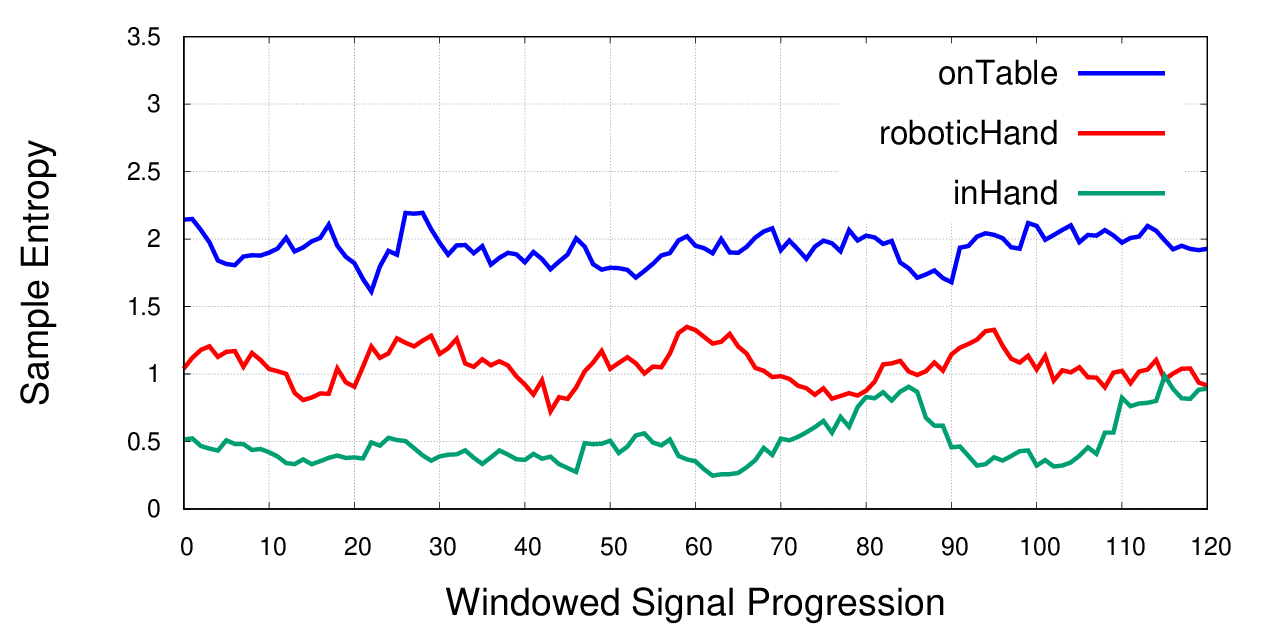}
         \caption{ Sample Entropy}
         \label{fig:sampleEntropy}
     \end{subfigure}
     \begin{subfigure}{0.333\textwidth}
         \centering         \includegraphics[width=\textwidth]{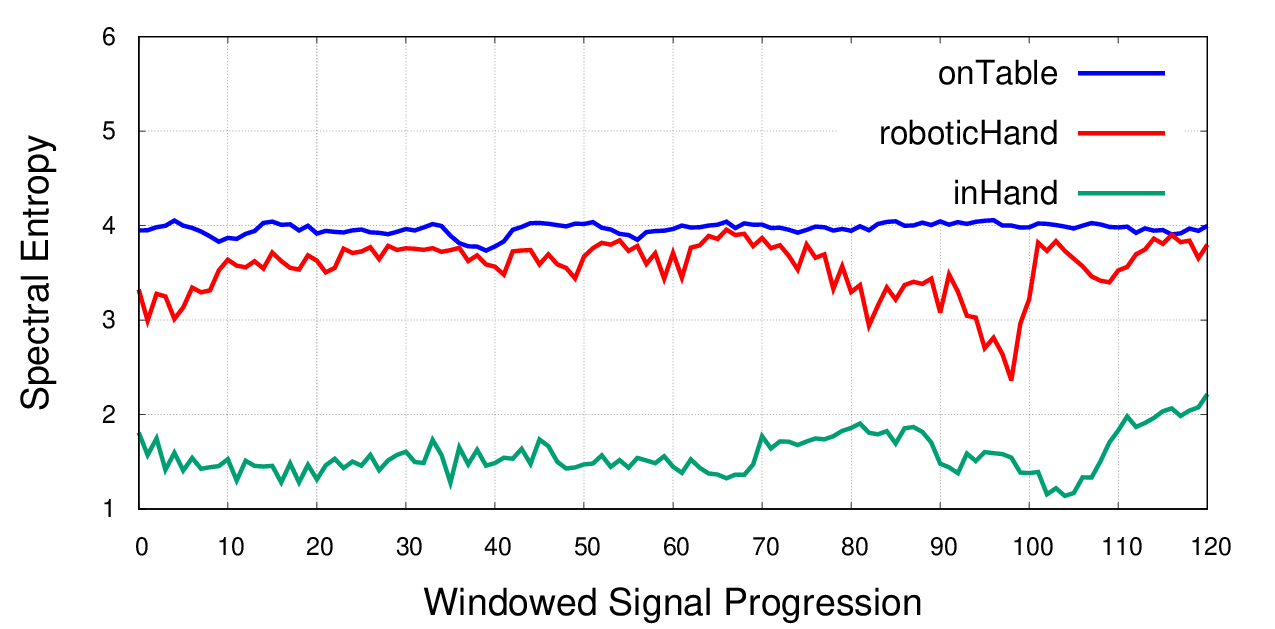}
         \caption{ Spectral Entropy}
         \label{fig:spectral_entropy}
     \end{subfigure}
     \begin{subfigure}{0.333\textwidth}
         \centering         \includegraphics[width=\textwidth]{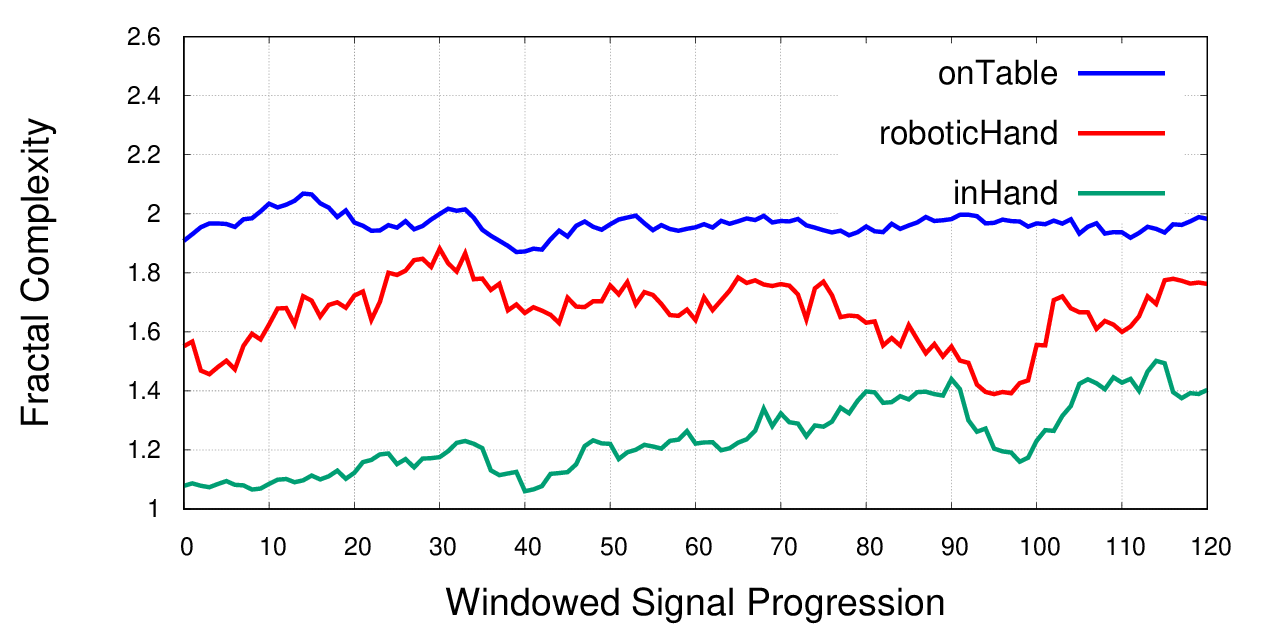}
         \caption{Fractal Complexity}
         \label{fig:fractalComplexity}
     \end{subfigure}
       \caption{Temporal evolution of representative motion complexity features for human, static (on-table), and programmable micro-motion adversarial signals. While raw signals may appear visually similar, the extracted features exhibit differences in variability and structural irregularity across conditions, illustrating their discriminative behavior in feature space.}
        \label{fig:features}
\end{figure*}

These features, which are computed using standard formulations from nonlinear time-series analysis and spectral signal processing, capture complementary aspects of motion dynamics, including irregularity, frequency distribution, and structural complexity. As shown in Figure~\ref{fig:features}, human-generated motion exhibits higher variability and less regular structure compared to both static (on-table) and mechanically generated motion signals, even when the raw signals appear visually similar. This reflects the inherent stochasticity and neuromuscular micro-variations present in natural human interaction.
In contrast to raw signal representations, where differences may be subtle, the feature space provides a more discriminative view in which the separation between human and non-human motion becomes significantly clearer. This highlights the role of feature extraction in amplifying subtle but consistent differences in motion dynamics.

\textbf{Multi-Resolution Frequency Analysis.}
To capture motion structure at different temporal scales, the signal is analyzed within a low-frequency range of $0-45\,\mathrm{Hz}$, which is consistent with the dominant frequency content of human motion captured by inertial measurement units in wearable sensing systems~\cite{freq2012}. Within this range, A-Live constructs multiple sub-band representations that capture motion structure at different resolutions, enabling the characterization of fine-grained variations in movement dynamics while maintaining robustness to residual low-frequency trends that persist after the pre-processing stage.

Rather than relying on a single spectral view, A-Live aggregates information across these sub-band representations to improve robustness across users and devices. These frequency-domain features are computed on axis-aligned signals and are used alongside PCA-based projections, which operate on the same windowed segment in a separate representation space.

\textbf{Feature Composition.}
The final representation combines features extracted across both the original sensor space and the decorrelated representation space. Specifically, low-level temporal and spectral statistics are derived from band-filtered axis-aligned signals, while higher-order structural descriptors are computed in the PCA-transformed space to capture invariant motion patterns. This hybrid design allows A-Live to jointly model raw motion dynamics and geometry-invariant representations of device movement.
The final representation consists of a compact feature vector derived from a combination of:
(i) principal-component-based motion projections,
(ii) power spectral statistics across multiple frequency bands,
(iii) spectral shape descriptors (e.g., flatness, energy concentration, and dominant frequency characteristics), and
(iv) robustness-oriented ratios designed to reduce sensitivity to device-specific motion artifacts.

This structured representation enables reliable discrimination between human-generated and non-human interaction patterns using a lightweight classifier operating on extracted feature vectors. By avoiding end-to-end learning over raw sensor streams, the design supports low-latency inference on commodity mobile devices without requiring specialized hardware or additional sensing infrastructure.
To facilitate reproducibility and real-world validation, we developed accompanying Android implementations that demonstrate the full data acquisition and inference pipeline.

\subsection{Classification and Decision Function}
\label{sec:classifier}

Given the extracted feature representation, A-Live performs liveness inference using a lightweight tree-based classification model designed for efficient and stable decision-making on structured motion features. The classifier operates on the assumption that the discriminative power of the system primarily resides in the engineered feature space, rather than in complex end-to-end representation learning.

To this end, we adopt an ensemble-based decision model inspired by gradient-boosted tree architectures. The choice of a tree-based formulation is motivated by three key properties: (i) robustness to feature scaling and heterogeneous signal distributions, (ii) strong performance on tabular and heterogeneous feature vectors, and (iii) low-latency inference suitable for mobile deployment. While alternative classifiers such as linear models or kernel methods can also operate on the same feature space, tree-based models provide a favorable balance between accuracy and computational efficiency in our setting.

A-Live produces a binary decision output $f(\mathbf{x}) \in \{0,1\}$ corresponding to non-human and human interaction classes, respectively. The decision is computed over short temporal windows of one second, enabling real-time inference without requiring temporal aggregation over long sequences. In addition to the binary output, the classifier can optionally produce a confidence score reflecting the stability of the decision boundary under the learned ensemble, which is used for threshold calibration in deployment scenarios. Formally, the classifier computes a real-valued decision score $g(\mathbf{z}_k)$ over the feature vector, and produces a binary output via thresholding:
\begin{equation}
f(\mathbf{z}_k) = \mathbb{I}\big(g(\mathbf{z}_k) \geq \tau \big),
\end{equation}
where $\mathbb{I}(\cdot)$ is the indicator function and $\tau$ is a decision threshold that can be calibrated to balance false acceptance and false rejection rates.

Importantly, the overall system is designed to be model-agnostic in the sense that the feature representation is sufficiently expressive that alternative lightweight classifiers yield consistent performance trends. This indicates that the primary discriminative signal in A-Live is embedded in the micro-movement feature space derived from inertial measurements, rather than being contingent on the specific choice of learning algorithm.

\subsection{Implementation Considerations}
\label{sec:implementation}

A-Live is designed for real-time execution on commodity mobile devices without requiring specialized hardware acceleration. The dominant latency in the pipeline arises from the sensor acquisition stage, which operates over a 1-second window to ensure sufficient motion signal fidelity. In contrast, feature extraction and classification are computationally lightweight, with inference executed in the millisecond range across a range of Android and iOS devices.

The system operates on tri-axial inertial data sampled at device-supported rates (typically around $100\,\mathrm{Hz}$ or higher). While shorter acquisition windows are feasible, we empirically observe that reducing the window length introduces a trade-off between latency and classification reliability; this trade-off is explored in the evaluation section.

The feature representation consists of approximately 50 descriptors, enabling compact processing and low memory overhead. Classification is performed using a shallow ensemble of decision trees (approximately 100 trees with bounded depth), resulting in predictable and efficient inference with minimal branching complexity. The model executes in a single-threaded setting and does not rely on GPU acceleration.

Overall, the computational footprint of A-Live is small, making it suitable for continuous or on-demand deployment in resource-constrained environments. As discussed earlier, we developed publicly deployable Android implementations that serve as practical validation of the system and demonstrate real-time end-to-end performance on commodity devices.

\section{Evaluation}
\label{sec::evaluation}

\begin{table*}[!t]
\centering
\caption{Comparison of representative liveness detection approaches across key system-level characteristics.}
\label{tab:comparison}
\resizebox{\textwidth}{!}{
\begin{tabular}{l l c c c c c}
\toprule
\textbf{Category} & \textbf{Method} & \textbf{Passive} & \textbf{User Interaction} & \textbf{No Special HW} & \textbf{Robustness} & \textbf{Scalable} \\
\midrule

\multirow{5}{*}{Challenge-based}
 & IMU-assisted motion~\cite{imuLiveness} & \xmark & High & \cmark & Medium & \cmark \\
 & Voice challenge~\cite{void} & \xmark & High & \cmark & Medium & \cmark \\
  & CAPTCHA~\cite{captcha} & \xmark & High & \cmark & Low & \cmark \\
 & Blink~\cite{blink} & \xmark & High & \cmark & Medium & \cmark \\
 & Gesture~\cite{livenessSmile} & \xmark & High & \cmark & Medium & \cmark \\
\midrule

\multirow{6}{*}{Appearance-based}
 & Texture-based~\cite{maata} & \cmark & None & \cmark & Low & \cmark \\
 & Deep learning~\cite{Liu2018} & \cmark & None & \cmark & Medium & \cmark \\
 & Reflection-based~\cite{Tan2010} & \cmark & None & \cmark & Medium & \cmark \\
 & Depth-based~\cite{Atoum2017} & \cmark & None & \xmark & Medium & \xmark \\
 & Color-variation rPPG~\cite{Verkruysse2008} & \cmark & Low & \cmark & Low & \cmark \\
 & BSS-based rPPG ~\cite{McDuff2010} & \cmark & Low & \cmark & Low & \cmark \\
\midrule

\multirow{7}{*}{Physiological}
 & ECG~\cite{Odinaka2012ECG} & \xmark & Medium & \xmark & High & \xmark \\
 & EMG~\cite{emg} & \xmark & Medium & \xmark & High & \xmark \\
 & Finger pulse~\cite{livenessFinger} & \xmark & Medium & \xmark & High & \xmark \\
 & Iris (Worldcoin)~\cite{worldcoinWhitePaper} & \xmark & Low & \xmark & High & \xmark \\
 & BCG~\cite{bcg} & \cmark & Low & \cmark & Medium & \cmark \\
& Iris (camera)~\cite{irisLiveness} & \cmark & Low & \cmark & Medium & \cmark \\\cline{2-7}
 & \textbf{A-Live (this work)} & \textbf{\cmark} & \textbf{None} & \textbf{\cmark} & \textbf{High} & \textbf{\cmark} \\

\bottomrule
\end{tabular}
}
\end{table*}

Evaluating a fundamentally novel liveness detection paradigm introduces unique challenges due to the absence of directly comparable prior work operating under the same signal modality and threat model. As discussed in Section~\ref{sec::relatedWork}, existing approaches span heterogeneous categories with differing assumptions, sensing requirements, and interaction models. Consequently, we adopt a multi-dimensional evaluation methodology that examines (i) empirical performance under realistic and adversarial conditions, (ii) robustness across heterogeneous devices and sensing configurations, and (iii) comparative positioning with respect to established classes of liveness detection systems.

Overall, our evaluation spans three complementary axes: large-scale passive spoofing via commercial device farms, natural human interactions, and actively engineered physical micro-motion attacks. This progression allows us to assess both statistical performance at scale and resilience under increasingly powerful adversarial capabilities.
To reflect these objectives, we structure the evaluation as follows. We begin with a comparative analysis situating A-Live within the broader landscape of liveness detection approaches. We then describe the experimental setup and evaluation environment under which A-Live is deployed and exercised, followed by the evaluation protocol and metrics. Next, we examine robustness against physically realizable micro-motion attacks. Finally, we present quantitative and qualitative results across all evaluation settings.

\subsection{Comparative Analysis with Existing Approaches}

To contextualize A-Live within the broader liveness detection landscape, we present a structured comparison against representative prior work in Section~\ref{sec::relatedWork}. Given the lack of consistent experimental conditions across prior studies, we focus on system-level properties rather than direct performance comparisons. Specifically, we evaluate sensing modality, interaction requirements, hardware dependency, robustness, and deployment scalability.

Table~\ref{tab:comparison} summarizes this comparison across challenge-based, appearance-based, and physiological-based methods. As shown, existing approaches typically trade off between usability, robustness, and scalability. In contrast, A-Live leverages neuromuscular micro-movements to achieve passive, real-time operation with strong robustness and high scalability, without requiring specialized hardware or explicit user interaction.

\subsection{Experimental Setup and System Evaluation}
Our evaluation is conducted in an \emph{in-situ} manner, where A-Live is executed directly on commodity devices and operates on live IMU sensor streams rather than pre-collected datasets. Specifically, the system is deployed on Android and iOS devices spanning a broad hardware spectrum.

The Android evaluation covers 61 device models from nine manufacturers, including Samsung, Google, OnePlus, Oppo, Xiaomi, Vivo, Motorola, Realme, and Huawei, with sensor sampling rates ranging from $199\,\mathrm{Hz}$ to $513\,\mathrm{Hz}$. The iOS evaluation includes 40 device models spanning multiple iPhone and iPad generations, providing inertial measurements at a system-defined sampling rate of approximately $100\,\mathrm{Hz}$, as exposed by the iOS motion sensor APIs. This difference reflects platform-level variations in sensor API design and hardware configurations rather than any limitation of the proposed method. A-Live is designed to operate independently of sampling rate variations through fixed-duration windowing and time-normalized feature extraction, without requiring resampling or interpolation across devices.

Non-human interactions are generated through commercial device farm infrastructures, including AWS Device Farm, SmartBear BitBar, SauceLabs, and BrowserStack. In these experiments, A-Live is executed remotely on target devices, processing live sensor streams produced by automated interaction scripts, thereby forming the spoof class in a fully operational setting.

For human evaluation, the system is deployed on devices operated by participants performing natural phone interactions in real-world conditions, forming the live class. In both cases, A-Live processes continuous IMU streams in a streaming fashion using 1-second windows with 1-second stride, and reports decisions per window without offline buffering or dataset construction.
For device farm and automated evaluation, 1000 independent interaction sessions are executed per device model under identical configuration settings.

\subsection{Evaluation Protocol and Metrics}

We formulate the task as binary classification between live (human) and spoof (non-human) interactions. Let TP and TN denote correct classifications of live and spoof samples, respectively, while FP and FN denote misclassifications. Formally, the False Acceptance Rate (FAR) and False Rejection Rate (FRR) are defined as:
\begin{equation}
\text{FAR} = \frac{\text{FP}}{\text{FP} + \text{TN}}, \qquad
\text{FRR} = \frac{\text{FN}}{\text{FN} + \text{TP}}.
\end{equation}

We focus on FAR and FRR as the primary evaluation metrics, as they directly capture the security–usability trade-off at a fixed operating point in liveness detection systems. Our evaluation is deployment-oriented, spanning large-scale device-farm automation, real human interactions, and controlled physical micro-motion attacks (Section~\ref{sec:robustness}). In this setting, the system operates using a fixed decision threshold, reflecting realistic runtime conditions.

The decision threshold was determined during a separate development phase using disjoint data and then held constant throughout all large-scale evaluations. This design avoids post-hoc threshold tuning on evaluation data and ensures that reported results reflect unbiased deployment behavior. While threshold-sweeping analyses such as ROC or DET curves are commonly used to characterize classifier performance across operating regimes and were used during development for threshold selection, they are not the focus of our large-scale evaluation. Instead, we report performance at the fixed operating point used during system execution, aligning directly with real-world deployment conditions.

\subsection{Robustness to Physical Micro-Motion Attacks}
\label{sec:robustness}

Beyond passive and human evaluations, we introduce a programmable micro-motion attack platform that models active physical adversaries capable of generating engineered motions directly at the device level. This setup addresses a key limitation of device-farm-based evaluation, which does not capture adversarially controlled motion dynamics with explicit mechanical structure.
As shown in Figure~\ref{fig:rawdata}, while coarse motion patterns can be mechanically reproduced and appear visually similar, the fine-grained structure associated with human neuromuscular micro-movements is not readily discernible to the naked eye. However, these subtle dynamics are systematically captured by the engineered feature set of A-Live and become distinguishable in the resulting feature space, as illustrated in Figure~\ref{fig:features}.

The platform consists of a smartphone mounting system mechanically coupled with multiple independently controllable electric motors. Each motor is driven via a dedicated power supply and stepless speed controller, enabling fine-grained control over actuation frequency and intensity. The number of active motors is configurable, and their spatial placement relative to the mounting structure is a critical design parameter affecting the resulting motion dynamics.
To expand the adversarial space, we introduce several degrees of freedom in the attack configuration. Circular as well as intentionally eccentric masses can be attached to motor shafts, while in certain configurations motors operate without any eccentric load, resulting in distinct motion regimes. We further explore all combinations of motor counts, placement configurations, mass configurations, actuation frequencies, and synchronization modes. In particular, both synchronized and asynchronous motor actuation are evaluated, as they play a key role in producing motion patterns that approximate human-like variability.
Figure~\ref{fig:roboticHand} illustrates the programmable motorized platform used to generate adversarial micro-motion patterns, including both the system schematic and one representative physical instantiation among the diverse configurations evaluated.

\begin{figure*}
    \centering

    \begin{subfigure}{\columnwidth}
        \centering
        \includegraphics[width=0.65\textwidth]{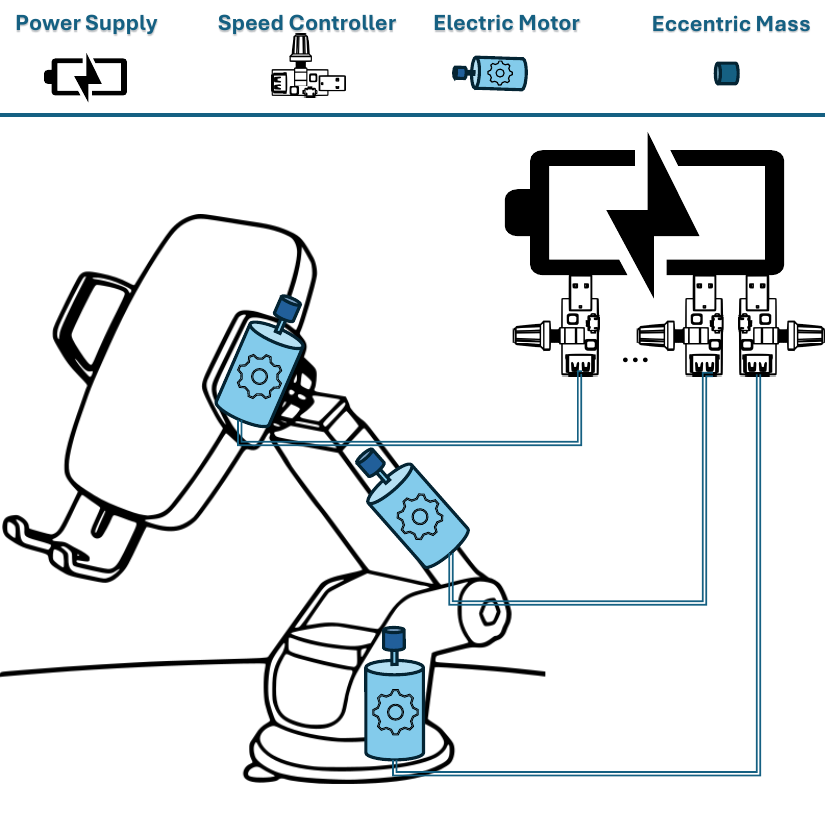}
        \caption{System schematic}
        \label{fig:roboticHand_schematic}
    \end{subfigure}
    \hfill
    \begin{subfigure}{\columnwidth}
        \centering
        \includegraphics[width=0.65\textwidth]{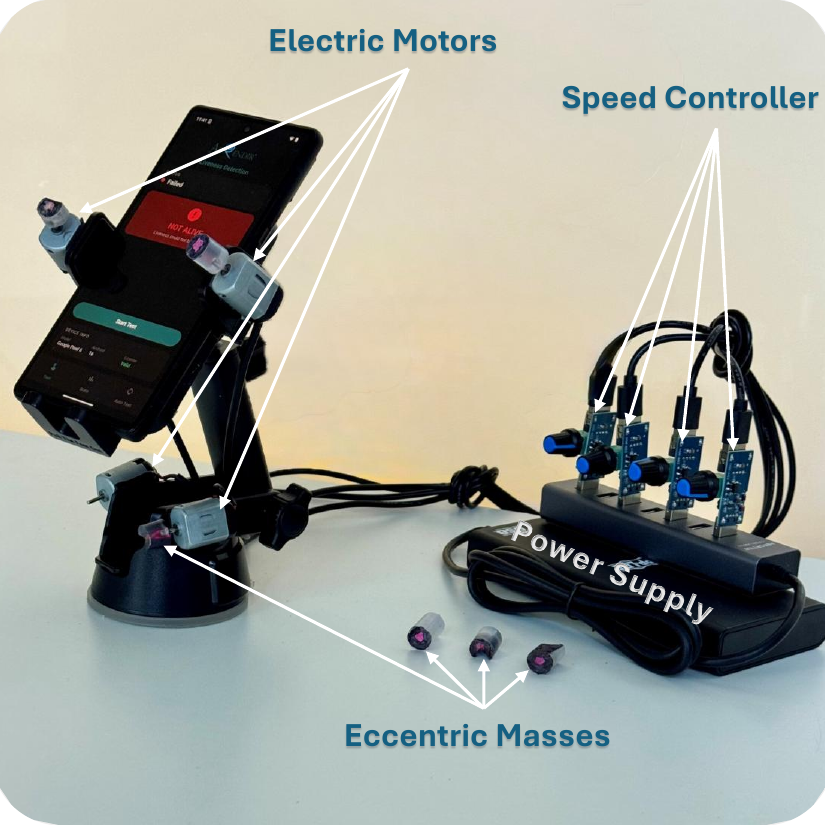}
        \caption{Physical prototype}
        \label{fig:roboticHand_pic}
    \end{subfigure}

    \caption{Programmable adversarial motorized platform for generating controlled micro-motion perturbations.}
    \label{fig:roboticHand}
\end{figure*}

Across all evaluated configurations, the proposed system consistently rejects all attacks, achieving a 0\% attack success rate. This result demonstrates robustness under a substantially stronger physical adversarial threat model than passive evaluation alone, and validates the effectiveness of neuromuscular micro-movement signals in distinguishing live human interactions from engineered mechanical motion.

\subsection{Results}
\label{sec:results}

We evaluate A-Live under large-scale real-world and adversarial deployment conditions across heterogeneous Android and iOS devices. The results are reported in terms of False Acceptance Rate (FAR), capturing security failures (spoof accepted as live), and False Rejection Rate (FRR), capturing usability impact (live rejected as spoof). The detailed evaluation results across platforms are summarized in Table~\ref{tab:results}, and discussed below.

\begin{table}[t]
\centering
\caption{Evaluation results across Android and iOS platforms under spoof and live conditions.}
\label{tab:results}
\setlength{\tabcolsep}{4pt}
\begin{tabular}{lcccccc}
\toprule
\textbf{Platform} & \textbf{Spoof Trials} & \textbf{FP} & \textbf{FAR (\%)} & \textbf{Live Trials} & \textbf{FN} & \textbf{FRR (\%)} \\
\midrule
Android & 61{,}000 & 9 & 0.015 & 3{,}000 & 13 & 0.43 \\
iOS     & 40{,}000 & 0 & 0.000 & 2{,}500 & 8  & 0.32 \\
\bottomrule
\end{tabular}
\end{table}

Across Android device-farm evaluations, we execute 61,000 spoof trials, resulting in 9 false acceptances, corresponding to a FAR of 0.015\%. On iOS devices, 40,000 spoof trials produce zero false acceptances, yielding an observed FAR of 0\% under this configuration.

To evaluate usability under genuine interaction, we conduct 3,000 live trials on Android devices and 2,500 live trials on iOS devices. The system produces 13 false rejections on Android (FRR = 0.43\%) and 8 false rejections on iOS (FRR = 0.32\%), indicating consistently low rejection rates across platforms.

Overall, A-Live exhibits near-zero false acceptance rates under large-scale spoofing and consistently low false rejection rates under genuine usage across heterogeneous mobile platforms. Notably, the system maintains robust performance even under high-volume evaluation, where small error rates translate into significant absolute security risk at scale. These results support the claim that neuromuscular micro-movement signatures provide a reliable discriminative signal for separating live human interactions from non-human or adversarially generated inputs under real-world conditions.

\section{Conclusion}
\label{sec::conclusion}

In this work, we presented A-Live, a passive liveness detection framework that leverages neuromuscular micro-movements captured through commodity IMU sensors. Unlike most prior approaches, which trade off usability, deployability, or robustness, A-Live operates transparently and continuously without requiring explicit user interaction, specialized hardware, or appearance-based sensing.
Our key insight is that fine-grained neuromuscular activity induces structured stochastic variations in inertial signals that remain difficult to replicate through non-human or adversarial mechanisms. We show that these micro-motion signatures, that often treated as noise, provide a reliable and discriminative signal for distinguishing live human interactions from synthetic or mechanically generated motion.
Through large-scale evaluation across heterogeneous Android and iOS devices, commercial device farms, real human interactions, and a programmable micro-motion attack platform, A-Live achieves near-zero false acceptance rates and consistently low false rejection rates, while maintaining robustness under active physical adversarial conditions.

Beyond liveness detection, our findings suggest that neuromuscular micro-movement signals reflect underlying neurophysiological processes and may serve as a novel biometric primitive for human presence verification. This opens the possibility of continuous and passive authentication systems that leverage intrinsic neuromuscular activity as a stable and distinctive signal, with the potential to complement, extend, and in certain settings reduce reliance on existing biometric modalities.
Separately, a key practical limitation of A-Live is its reliance on IMU-equipped devices, which may not be directly available in desktop or laptop environments. An important direction for future work is the design of secure cross-device liveness attestation mechanisms, where a trusted mobile device performs liveness detection and generates a cryptographically bound, tamper-resistant, and non-forgeable attestation that can be securely transmitted to and verified by other platforms through a secure and low-latency transport layer.
Overall, A-Live demonstrates that meaningful physiological structure exists in low-level inertial sensor data, and that such signals can be leveraged to build practical, scalable, and hardware-efficient security primitives for next-generation human liveness detection systems.

\nocite{*}
\bibliographystyle{IEEEtran}
\bibliography{References}

\end{document}